\documentclass[10pt,letterpaper,twocolumn]{article} 

\usepackage{ol2}
\usepackage[draft]{hyperref}
\usepackage{amsbsy,amsmath,amssymb}
\usepackage{setspace}
\usepackage{upgreek}
\usepackage{graphicx}

\begin{document}

\twocolumn[ 

\title{Spin Hall effect of light in metallic reflection}


\author{N. Hermosa,$^{1,*}$ A.M. Nugrowati,$^1$ A. Aiello,$^{2,3}$ and J.P. Woerdman$^1$}
\address{$^1$Huygens Laboratory, Leiden University, P.O. Box 9504, 2300 RA Leiden, The Netherlands\\$^2$Max Planck Institute for the Science of Light, G\"{u}nter-Scharowsky-Stra\ss e 1/Bau 24, 91058 Erlangen, Germany\\
$^3$Institute for Optics, Information and Photonics, Universit\"{a}t Erlangen-N\"{u}rnberg, Staudtstr. 7/B2, 91058 Erlangen, Germany\\
$^*$Corresponding author: hermosa@physics.leidenuniv.nl
}

\begin{abstract}We report the first measurement of the Spin Hall Effect of Light (SHEL) on an air-metal interface. The SHEL is a polarization-dependent \textit{out-of-plane} shift on the reflected beam. For the case of \textit{metallic} reflection with a linearly polarized incident light, \textit{both} the spatial and angular variants of the shift are observed and are maximum for $-45^\circ/45^\circ$ polarization, but zero for pure $s$- and $p$-polarization. For an incoming beam with circular polarization states however, only the \textit{spatial} out-of-plane shift is present. 
\end{abstract}


 ] 

\noindent The Spin Hall Effect of Light (SHEL) is the photonic analog of the Spin Hall Effect in solid state physics in which the spin of the particles are replaced by the spin of photons (i.e. polarization) and the electric potential gradient by the refractive index gradient \cite{Onoda:PRL2004, Bliokh:PRL2006, Hosten:Science2008}.  The SHEL appears as a very small but detectable polarization dependent \textit{out-of-plane} (namely, transverse to the plane of incidence) displacement of the reflected beam at a dielectric interface relative to the geometric-optics prediction.  Introduced in~\cite{Onoda:PRL2004} as a transport phenomenon, the effect was in fact first theoretically derived by Fedorov in 1955 for the case of total internal reflection in glass \cite{Fedorov:1955}. Its experimental verification was done by Imbert in 1972 \cite{Imbert:PRD1972}; hence the SHEL is also known as the Imbert-Fedorov (IF) effect. Recently, there has been a renewed interest in the SHEL both regarding its theoretical understanding \cite{Onoda:PRL2004, Bliokh:PRL2006, Aiello:OptLett2008, Fedoseyev:OptCommun2009, Luo:PRA2009} and its potential for metrology applications. A considerable amount of experimental work has been reported for an air-dielectric interface \cite{Hosten:Science2008, Pillon:04, Haefner:PRL2009, Qin:OptLetter2009, Merano:PRA10, Qin:OpEx10} and for an air-semiconductor interface \cite{Menard:PRB2010}. 
\begin{figure}[htbp]
\begin{center}
\includegraphics[width=7.5cm]{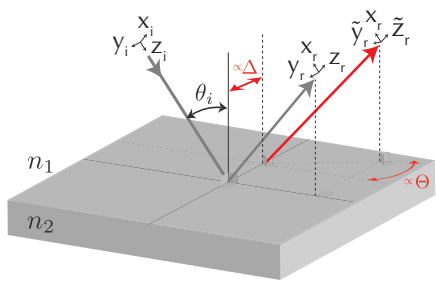}
\end{center}
\caption{\label{fig:cartoon}(Color online) An illustration of \textit{out-of-plane} spatial ($\propto\Delta$) and angular ($\propto\Theta$) shifts of a beam incident at an angle $\theta_i$,  upon reflection at an interface of two media.}
\end{figure}

The SHEL is part of the diffraction correction on the path of a bounded beam upon its reflection or refraction, which cannot be exactly described by the geometrical-optics (Snell's law and Fresnel formulas) alone \cite{Bliokh:PRL2006}. This small correction also includes the \textit{in-plane} shift \cite{Goos:AnnPhys1947}, known as Goos-H\"{a}nchen (GH) shift.  For both the SHEL and GH cases, there exist two variants of shift namely the spatial and angular shift; the latter being enhanced by propagation \cite{Aiello:OptLett2008}. Figure~\ref{fig:cartoon} illustrates the spatial and angular shift for the SHEL. 

In this Letter, we report for the first time, experimental measurements of the SHEL on an air-metal interface for different states of polarizations. We are interested in the SHEL on an air-metallic interface because metals have complex refractive indices, as compared to the purely real refractive indices of dielectric media. 

Theoretically, a consequence of using metallic reflection has been reported previously~\cite{Aiello:PRA2009}, demonstrating that the spatial and angular shifts can be described by a common formalism; experimental evidence for this was provided for the GH case but not for the SHEL. Several studies have in fact shown that GH shifts is effectively a scalar effect in the sense that $s$- and $p$-polarized light undergo individual (uncoupled) GH effects~\cite{Merano:NatPhoton2009, Aiello:OptLett2008, Merano:OptLett2010}. The SHEL on the other hand, requires simultaneous $s$- and $p$-component (see Eq.~\ref{eqs:shift}), otherwise a transverse shift cannot occur due to symmetry reasons.

To calculate the out-of-plane shift of a beam with finite transverse extent, one can use either the law of conservation of angular momentum \cite{Onoda:PRL2004, Bliokh:PRL2006, Hosten:Science2008, Fedoseyev:OptCommun2009} or more directly, using angular spectrum decomposition \cite{CFLi:PhysRevA2007, Aiello:OptLett2008}. We used the latter method to derive the equations below. The incident and reflected beams are assumed to be Gaussian; they are decomposed into plane wave components, and the Fresnel reflection coefficients are applied for both the $s$- and $p$-component of the waves, respectively. The shift is then obtained in the paraxial approximation by integrating over the reflected plane waves.  

Following the notations described in Eq. (5) of~\cite{Aiello:OptLett2008}, we express the dimensionless \textit{out-of-plane} shift as:
\begin{equation}\label{eq:TotalShift}
Y_r=\Delta+Z_r\dfrac{\Theta}{\Lambda},
\end{equation}
where $Y_r=k_0y_r$, $Z_r=k_0z_r$, and $\Lambda=k_0(k_0w_0^2/2)=2/\theta_0^2$ with $k_0$ the wavenumber of beam center, $w_0$ the beam waist, and $\theta_0$ the opening angle of the incoming beam. Eq.~(\ref{eq:TotalShift}) yields the total out-of plane shift $y_r$ that consists of two parameters with measurable units: the spatial shift $\Delta/k_0$ as a dimensional length and the angular shift $\Theta/\Lambda$ in radian dimension, respectively. Variables $\Delta$ and $\Theta$ are expressed as
\begin{subequations} \label{eqs:shift}
\begin{align}
\label{eq:spatial}
\begin{split}
\Delta &= -\dfrac{a_p a_s \cot \theta_i}{R_p^2 a_p^2+R_s^2 a_s^2}\\
&\left[\left(R^2_p+R^2_s\right)\sin\eta+2R_p R_s \sin\left(\eta-\varphi_p+\varphi_s\right)\right],
\end{split}\\
\label{eq:angular}
\Theta &= \dfrac{a_p a_s \cot\theta_i}{R_p^2 a_p^2+R_s^2 a_s^2}\left[\left(R^2_p-R^2_s\right)\cos\eta\right],
\end{align}
\end{subequations}
with $r_{s/p}=R_{s/p}\exp(i\varphi_{s/p})$ the Fresnel reflection coefficient evaluated at incident angle $\theta_i$ and $a_{s/p}$ the electric field components for perpendicular and parallel directions, respectively. The phase shift between these two components is given by $\eta$. 

\begin{figure}[htbp]
\begin{center}
\includegraphics[width=8.3cm]{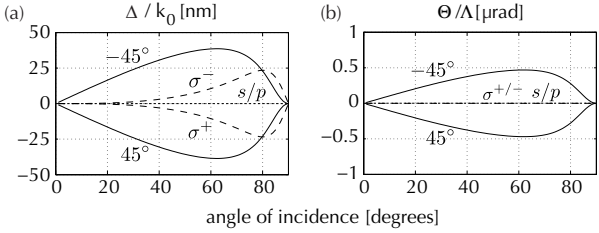}
\end{center}
\caption{\label{fig:theory}Theoretical curves for (a) spatial and (b) angular shifts of linear (solid lines: $-45^\circ/45^\circ$, dotted lines: s/p) and circularly (dashed lines:  $\sigma^{-/+}$) polarized light as a function of incident angles, calculated for gold at $\lambda=826$~nm, $n=1.88+i5.39$ \cite{Palik:1985}.}
\end{figure}

For a somewhat lossy surface such as gold, the difference in the phase acquired by the $s$- and $p$-component of the waves after reflection varies gradually with angle of incidence, between 0 or $\pi$  \cite{Born:1999}. The theoretical curve for various polarization states of the spatial $\Delta/k_0$ and angular $\Theta/\Lambda$ shifts is shown in Fig.~\ref{fig:theory}. A key distinction, as compared to the case of dielectric reflection, is that these two shifts can coexist, due to the finite losses of the metal (Au) \cite{Aiello:PRA2009}. For linearly but oblique polarized light both spatial and angular shifts contribute to the measurable beam shift $y_r$. The shifts are maximum at $\pm45^\circ$ polarization angle and becomes zero for $s$- and $p$-polarized light. Only the spatial shift occurs when using circularly polarized ($\sigma^{-/+}$) light.  

\begin{figure}[htbp]
\begin{center}
\includegraphics[width=8.3cm]{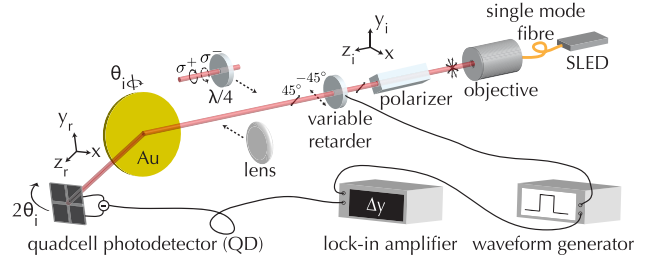}
\end{center}
\caption{\label{fig:setup}(Color online) The optical set-up to measure the polarization-differential shift as a function of incident angle $\theta_i$. See text for details.}
\end{figure}

Upon reflection on metal, a linearly polarized light will emerge elliptically polarized \cite{Born:1999} due to the different phase acquired by the $s$- and $p$-component of the waves. This makes post-selection scheme necessary to observe the SHEL via a weak measurement \cite{Hosten:Science2008,Qin:OptLetter2009} impractical. We employ a different scheme, shown in Fig.~\ref{fig:setup}.  We use an 826 nm superluminescent light-emitting diode (SLED) which is spatially filtered by a single-mode optical fiber to yield a TEM$_{\textrm{00}}$ mode, collimated at a beam waist of $w_0=690\mu m$ and polarized by a Glan polarizer.  To measure polarization-differential shifts, we switch between orthogonal polarization states ($-45^\circ/45^\circ$ or $\sigma^-/\sigma^+$) at 2.5Hz, with a liquid-crystal variable retarder (LCVR, Meadowlark).  To create circularly polarized right, a QWP is inserted. Our sample is a planar (Wyko optical profiler gives $0.8$~nm rms roughness) $200$~nm thick Au film that is deposited on a chromium film-coated Duran ceramic glass (diameter = 10 cm, surface flateness = $\lambda$/20). 
The polarization-differential shifts of the reflected beam are detected by a calibrated quadcell photodetector (QD, model 2901/2921 NewFocus).  To increase the opening angle of the beam, we insert a lens before the beam hits the sample.  We determine the contribution of the angular shift from the measured total shift by varying the position of the detector with respect to the waist of the focused beam.

With the use of a lock-in amplifier (EG\&G 5210), all measurements are performed by synchronously measuring the relative transverse position (along the $y_r$-axis) of the beam while switching polarizations with the LCVR. We obtained the direction (positive vs negative) of the transverse shift of the beam by noting the phase of the lock-in amplifier. Due to the different intensities between $s$- and $p$-polarization, the signal being detected by the detector and read by the lock-in amplifier needs a correction, where we have generalized the recipe in \cite{Merano:OpEx2007} for any state of polarization, into:
\begin{equation}\label{eq:LockIn}
\delta y_r=\dfrac{\delta U}{C \Sigma_1} - \dfrac{U_{1}}{C\Sigma_1}\dfrac {(\Sigma_2-\Sigma_1)}{\Sigma_1},
\end{equation}
with $\delta y_r$ is the transverse shift in length units,  $\delta U$ the measured voltage difference read by the lock-in amplifier, and $C$ the calibration constant. Note that the subscripts \{$_{1,2}$\} are assigned to the switching polarization states in our experiments. The second term on the right hand side of Eq.~(\ref{eq:LockIn}) is the necessary correction to the read signal. It is minimum when $U_1$ is zero, i.e. the reflected beam with one of the switching polarization states is centered to the QD. Both  $U_1$ and the total intensity $\Sigma$ are measured by a voltmeter (HP 34401A). 

\begin{figure}[htbp]
\begin{center}
\includegraphics[width=8.3cm]{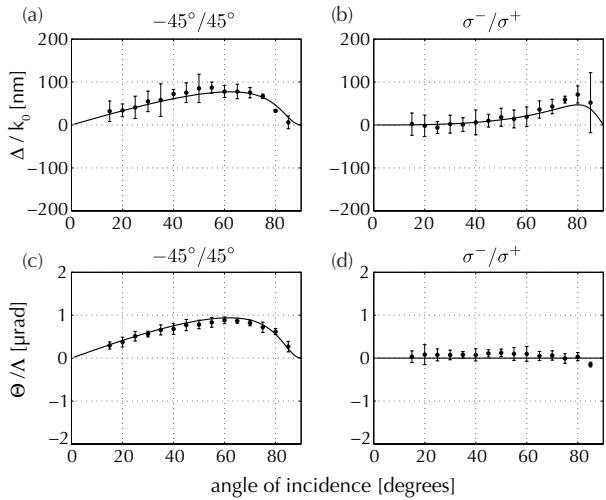}
\end{center}
\caption{\label{fig:data}Measured (top row) spatial and (bottom row) angular polarization-differential shifts between (a,c) $-45^\circ/45^\circ$ and (b,d) $\sigma^{-/+}$ polarized light, as a function of angle of incidence. Solid lines are theoretical curves, dots with error bars represent data with its standard deviation ($3\sigma$).}
\end{figure}

Figure~\ref{fig:data} shows our experimental results for the polarization-differential shifts as a function of the angle of incidence together with theoretical predictions based on Eqs.~(\ref{eq:TotalShift}-\ref{eqs:shift}). The data agrees with the theory without the use of fit parameters. The experimental error is of the order of 30 nm (to be compared with the 1 nm error in the weak-measurement method for dielectrics \cite{Hosten:Science2008, Qin:OptLetter2009}). 

In Fig.~\ref{fig:data}(a) and \ref{fig:data}(c), the angle of polarization is switched between $-45^\circ/45^\circ$. \textit{Both} the out-of-plane spatial and angular polarization-differential shifts are observed and peak at an incident angle of $65^\circ$. The measurement of the spatial Fig.~\ref{fig:data}(b) and angular Fig.~\ref{fig:data}(d) polarization-differential shifts between right and left circularly polarized light ($\sigma^{-/+}$) shows that only the \textit{spatial} out-of-plane shift is present.  We have also measured the out-of-plane shift for pure $s$- and $p$-polarization (plots not shown), and the results agree with predicted zero values within the limits of uncertainty.

In agreement with theory described in Eqs.~(\ref{eq:TotalShift}-\ref{eqs:shift}) above, our measurements of the SHEL in metal reflection do not show an existence of a backward energy flow, unlike in the GH case where the \textit{in-plane} shift becomes negative for gold.  In the GH shift, the energy flow of the evanescent field parallel to the interface changes sign when the sign of the permittivity $\epsilon$ of the reflecting medium changes~\cite{Leung:OptCommun2008, Lai:PRE2000}. This argument does not hold with the SHEL in metallic reflection as the shift is perpendicular to the incoming wave vector and therefore the sign is independent of $\epsilon$.     

In summary, we have demonstrated the SHEL on an air-gold interface for different polarizations. The SHEL is specially interesting in applications where minute shift can play an important role, e.g. in metrology. On a more fundamental level, our measurements add to the understanding of how the conservation law of angular momentum plays a role when light undergoes reflection\cite{Bliokh:PRL2006, Hugrass:JMO1990}.

This work is supported by the Foundation for Fundamental Research of Matter (FOM) and the European Union within FET Open-FP7 ICT as part of STREP Program 255914 Phorbitech.

\pagebreak

\section*{Informational Fourth Page}


\begin{thebibliography}{10}
\newcommand{\enquote}[1]{``#1''}

\bibitem{Onoda:PRL2004}
M.~Onoda, S.~Murakami, and N.~Nagaosa, \enquote{Hall effect of light,} Phys.
  Rev. Lett. \textbf{93}, 083901 (2004).

\bibitem{Bliokh:PRL2006}
K.~Y. Bliokh and Y.~P. Bliokh, \enquote{Conservation of angular momentum,
  transverse shift, and spin hall effect in reflection and refraction of an
  electromagnetic wave packet,} Phys. Rev. Lett. \textbf{96}, 073903 (2006).

\bibitem{Hosten:Science2008}
O.~Hosten and P.~Kwiat, \enquote{Observation of the spin hall effect of light
  via weak measurements,} Science \textbf{319}, 787--790 (2008).

\bibitem{Fedorov:1955}
F.~I. Fedorov, \enquote{K teorii polnogo otrazheniya,} Dokl. Akad. Nauk SSR
  \textbf{105}, 465 (1955).

\bibitem{Imbert:PRD1972}
C.~Imbert, \enquote{Calculation and experimental proof of the transverse shift
  induced by total internal reflection of a circularly polarized light beam,}
  Phys. Rev. D \textbf{5}, 787--796 (1972).

\bibitem{Aiello:OptLett2008}
A.~Aiello and J.~P. Woerdman, \enquote{Role of beam propagation in
  goos--h\"{a}nchen and imbert--fedorov shifts,} Opt. Lett. \textbf{33},
  1437--1439 (2008).

\bibitem{Fedoseyev:OptCommun2009}
V.~Fedoseyev, \enquote{Conservation laws and angular transverse shifts of the
  reflected and transmitted light beams,} Opt. Commun. \textbf{282}, 1247 --
  1251 (2009).

\bibitem{Luo:PRA2009}
H.~Luo, S.~Wen, W.~Shu, Z.~Tang, Y.~Zou, and D.~Fan, \enquote{Spin hall effect
  of a light beam in left-handed materials,} Phys. Rev. A \textbf{80}, 043810
  (2009).

\bibitem{Pillon:04}
F.~Pillon, H.~Gilles, and S.~Girard, \enquote{Experimental observation of the
  imbert-fedorov transverse displacement after a single total reflection,}
  Appl. Opt. \textbf{43}, 1863--1869 (2004).

\bibitem{Haefner:PRL2009}
D.~Haefner, S.~Sukhov, and A.~Dogariu, \enquote{Spin hall effect of light in
  spherical geometry,} Phys. Rev. Lett. \textbf{102}, 123903 (2009).

\bibitem{Qin:OptLetter2009}
Y.~Qin, Y.~Li, H.~He, and Q.~Gong, \enquote{Measurement of spin hall effect of
  reflected light,} Opt. Lett. \textbf{34}, 2551--2553 (2009).

\bibitem{Merano:PRA10}
M.~Merano, N.~Hermosa, J.~P. Woerdman, and A.~Aiello, \enquote{How orbital
  angular momentum affects beam shifts in optical reflection,} Phys. Rev. A
  \textbf{82}, 023817 (2010).

\bibitem{Qin:OpEx10}
Y.~Qin, Y.~Li, X.~Feng, Z.~Liu, H.~He, Y.-F. Xiao, and Q.~Gong, \enquote{Spin
  hall effect of reflected light at the air-uniaxial crystal interface,} Opt.
  Express \textbf{18}, 16832--16839 (2010).

\bibitem{Menard:PRB2010}
J.-M. M\'enard, A.~E. Mattacchione, H.~M. van Driel, C.~Hautmann, and M.~Betz,
  \enquote{Ultrafast optical imaging of the spin hall effect of light in
  semiconductors,} Phys. Rev. B \textbf{82}, 045303 (2010).

\bibitem{Goos:AnnPhys1947}
F.~Goos and H.~H\"{a}nchen, \enquote{Ein neuer und fundamentaler versuch zur
  totalreflexion,} Annalen der Physik \textbf{436}, 333--346 (1947).

\bibitem{Aiello:PRA2009}
A.~Aiello, M.~Merano, and J.~P. Woerdman, \enquote{Duality between spatial and
  angular shift in optical reflection,} Phys. Rev. A \textbf{80}, 061801
  (2009).

\bibitem{Merano:NatPhoton2009}
M.~Merano, A.~Aiello, M.~van Exter, and J.~Woerdman, \enquote{Observing angular
  deviations in the specular reflection of a light beam,} Nat. Photon.
  \textbf{06} (2009).

\bibitem{Merano:OptLett2010}
M.~Merano, N.~Hermosa, A.~Aiello, and J.~P. Woerdman, \enquote{Demonstration of
  a quasi-scalar angular goos-h\"{a}nchen effect,} Opt. Lett. \textbf{35},
  3562--3564 (2010).

\bibitem{CFLi:PhysRevA2007}
C.-F. Li, \enquote{Unified theory for goos-h\"{a}nchen and imbert-fedorov
  effects,} Phys. Rev. A \textbf{76}, 013811 (2007).

\bibitem{Palik:1985}
E.~Palik, ed., \emph{Handbook of Optical Constants of Solids} (Academic Press,
  New York, 1985).

\bibitem{Born:1999}
M.~Born and E.~Wolf, \emph{Principles of Optics} (Cambridge University Press,
  Cambridge, 1999), 7th ed.

\bibitem{Merano:OpEx2007}
M.~Merano, A.~Aiello, G.~W. 't~Hooft, M.~P. van Exter, E.~R. Eliel, and J.~P.
  Woerdman, \enquote{Observation of goos-h\"{a}nchen shifts in metallic
  reflection,} Opt. Express \textbf{15}, 15928--15934 (2007).

\bibitem{Leung:OptCommun2008}
P.~Leung, C.~Chen, and H.-P. Chiang, \enquote{Addendum to "large negative
  goos-h\"{a}nchen shift at metal surfaces", [opt. comm. 276 (2007) 206],} Opt.
  Commun. \textbf{281}, 1312 -- 1313 (2008).

\bibitem{Lai:PRE2000}
H.~Lai, C.~Kwok, Y.~Loo, and B.~Xu, \enquote{Energy-flux pattern in the
  goos-h\"{a}nchen shift,} Phys. Rev. E \textbf{62}, 7330 (2000).

\bibitem{Hugrass:JMO1990}
W.~Hugrass, \enquote{Angular momentum balance on light reflection,} J. Mod.
  Optics \textbf{37}, 339--351 (1990).

\end{thebibliography}
\end{document}